# Influence of thermal and resonance neutron on fast neutron flux measurement by $^{239}$Pu fission chamber[*]

Lina Zeng[1,2], Qiang Wang[1,2], Lingli Song[1,2], Chun Zheng[1,2*]

[1] Institute of Nuclear Physics and Chemistry, China Academy of Engineering Physics, Mianyang 621900, China

[2] Key Laboratory of Neutron Physics, China Academy of Engineering Physics, Mianyang 621900, China

**Abstract:** The $^{239}$Pu fission chambers are widely used to measure fission spectrum neutron flux due to a flat response to fast neutrons. However, in the mean time the resonance and thermal neutrons can cause a significant influence on the measurement if they are moderated, which could be eliminated by using $^{10}$B and Cd covers. At a column enriched uranium fast neutron critical assembly, the fission reaction rates of $^{239}$Pu are measured as $1.791 \times 10^{-16}$, $2.350 \times 10^{-16}$ and $1.385 \times 10^{-15}$ per second for 15mm thick $^{10}$B cover, 0.5mm thick Cd cover, and no cover respectively. While the fission reaction rate of $^{239}$Pu is rapidly increased to $2.569 \times 10^{-14}$ for a 20mm thick polythene covering fission chamber. The average $^{239}$Pu fission cross-section of thermal and resonance neutrons is calculated to be 500b and 24.95b with the assumption of $1/v$ and $1/E$ spectra respectively, then thermal, resonance and fast neutron flux are achieved to be $2.30 \times 10^6$, $2.24 \times 10^6$ and $1.04 \times 10^8$ cm$^{-2}$ · s$^{-1}$.

## 1 Introduction

Neutron flux is a very important data for neutron source. Many studies focus their eyes on neutron flux [1, 2]. Fission chambers are widely used in online neutron flux measurement. It could be adopt for not only the fission neutron field but also the fusion neutron field [3]. Because the fission fragment signal in chamber with large reaction energy of about 200MeV is larger than those of the competing reaction and γ-rays, which facilitates the n-γ discriminating. Also there are many studies about the background of fission chamber[4]. Moreover, the fission cross-sections above 10keV of $^{239}$Pu change insignificantly, often by less than 5% when the neutron energy between 10keV and 5MeV. As a result, $^{239}$Pu fission chambers are usually adopted to measure fast neutron flux [5].

Large cross-section of a few thermal and resonance neutrons contributes most fission reaction in the chamber: the fission rate changes sharply within a small range of fractions of thermal and resonant neutrons. The average fission cross-section of $^{239}$Pu for fast neutrons is 1.72b. And the fission cross-section of $^{239}$Pu at thermal neutrons (at the energy of 0.025eV) is 744b, which is larger than 1.72b. Additionally, $^{237}$Np fission rate ratio relative to $^{235}$U fission rate per atom was measured to be 0.00439 to 0.0298 at Kyoto University Critical Assembly at five thermal cores [6]. Therefore the effect of thermal and resonance neutrons should be considered at measurements of fast neutron flux. The effect due to thermal and resonance neutrons is eliminated by using of $^{10}$B and Cd covers.

Neither moderator materials nor reflective materials currently exist in Godiva or Flattop type fast critical assembly [7]. Ionization chamber also has a lot of room for improvement. For example, we can change the structures, materials and so on to improve the ability of the ionization chamber [8]. The CERN built a new ionization chamber with fast timing properties for the measuring of cross-section which particularly focused on fast time response, the good background rejection capability, low-background and high detection efficiency [9]. In this work we choose the $^{239}$Pu fission chamber which is always used to measure fast neutron flux of critical assembly with Cd cover and $^{10}$B cover to absorb thermal neutrons and resonance neutrons. And small concentrations of $^{241}$Pu and $^{242}$Pu can be tolerated. In this report, we focus in particular on the improvement of neutron flux measurement accuracy by fission chamber.

---
[*] Supported by National Natural Science Foundation of China (91326109)
1) E-mail: zhengchun@caep.cn
2) E-mail: linazengnwpu@163.com

## 2 Experiment Measurements

### 2.1 Detector System

The measuring system consists of a $^{239}$Pu fission chamber, a preamplifier, a linear amplifier, and a multi-channel pulse analyzer. The fission chamber current pulse depends on specifications such as the filling gas pressure and the fission chamber geometry [10]. 0.168mg $^{239}$Pu is uniformly plated on a foil surface with a length of 10mm and a width of 13.8mm in the fission chamber filled with $4.0\times10^5$Pa Argon. The broad side rolled into a round is placed inside the copper tube with the outer diameter of 6mm and the inner 4.5mm, as shown in Fig. 1. The fission chamber was fixed at an identical position at the four measurements.

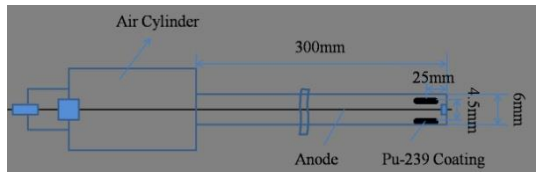

Fig.1. Schematic Diagram of the Fission Chamber

### 2.2 Neutron Spectrum of the Critical Assembly

The experiments were carried out in a $^{235}$U metal fast critical assembly. In the fast neutron energy region of the reactor, the neutron spectrum is slightly moderated fission spectrum. The calculated neutron spectrum of the critical assembly is shown in Fig.2. The average neutron energy is 1.42MeV.

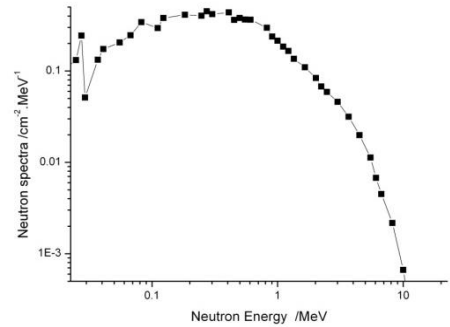

Fig.2. Neutron Spectrum of the U Fast Neutron Critical Assembly

For further analysis, the neutron spectrum is divided into three groups. The thermal neutron region, En<1eV, is assumed to be 1/v spectrum. The resonance neutron region, 1eV<En<10keV, is assumed to be 1/E spectrum. The fast neutron group, En>10keV, is assumed to be a slightly moderated fission spectrum. The calculated spectrum contains almost zero the thermal neutrons, 0.06% of the resonance neutrons, and 99.94% of the neutrons with energy above 10keV, in which has 3.3% of thermal neutrons with energy above 5MeV. The structural material around the assembly is not taken into consideration in the calculation of neutron spectrum. In fact, there are often a few thermal neutrons and resonance neutrons for the moderation of the structural material. These neutrons contribute to the most reactions in the $^{239}$Pu chamber despite a little fraction.

### 2.3 Experiments

The $^{239}$Pu fission chamber was fixed at the same position at four measurements, with the high voltage of 210V, and the critical assembly operating at 180W. The $^{239}$Pu fission fragments spectra were recorded by MCA. (a) the fission reaction rate of the naked $^{239}$Pu fission chamber was measured, (b) the reaction rate of the $^{239}$Pu fission chamber with a 0.5mm thick Cd cover was measured and (c) the reaction rate of the $^{239}$Pu fission chamber with $^{10}$B cover was measured. The length of $^{10}$B cover is 100mm, with the hollow length, the inner diameter and the outer diameter being 85mm, 10mm and 40mm respectively. There is a 0.5mm cadmium skin at the internal of the $^{10}$B cover. The effective cross-section of $^{239}$Pu with $^{10}$B cover is calculated as following:

$$\sigma_{\text{eff}}(E) = \sigma(E)e^{-\sigma_B(E)N} \quad (1)$$

Where: $\sigma_{\text{eff}}$ is the $^{239}$Pu effective cross-section, $\sigma(E)$ is the $^{239}$Pu fission cross-section, $\sigma_B(E)$ is the $^{10}$B cross-section and $N$ is the nuclei number of $^{10}$B per unit area.

In order to investigate the influence of thermal neutrons on the measurements, the fission reaction rate of the $^{239}$Pu fission chamber with a polythene cover was measured. The polythene cover is 19cm in length, 17cm in hollow length, 8cm in outer diameter and 4cm in inner.

## 3 Results and Discussions

The fission rate is calculated as following:

$$\dot{F} = N\phi\bar{\sigma}\eta \quad (2)$$

$\eta$ is the detection efficiency of the fission fragments where $\eta = 1$, $N$ is the number of the $^{239}$Pu, $\varphi$ is neutron flux, and $\bar{\sigma}$ is average cross-section.

### 3.1 Results of the Experiments

Fig.3 shows the measured fission fragments spectrum of the fission chamber. The double peaks in Fig.3 are corresponding to light and heavy fission fragments. The peaks of two fission fragments are clearly separated and α pulses are shown in the low energy region. A lower platform between the α and the fragments is shown in the enlarged view in Fig.3. The α and fission fragments can be well distinguished. The lower platform is extrapolated to take the fission fragments counts at the low-energy.

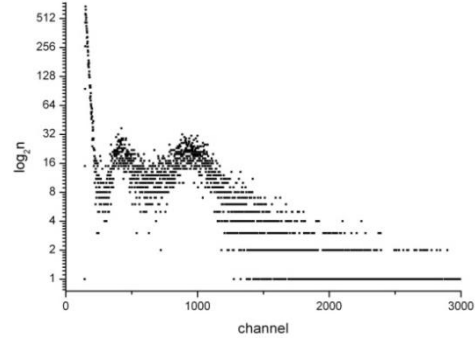

Fig.3. (a) Fission Fragment Spectrum

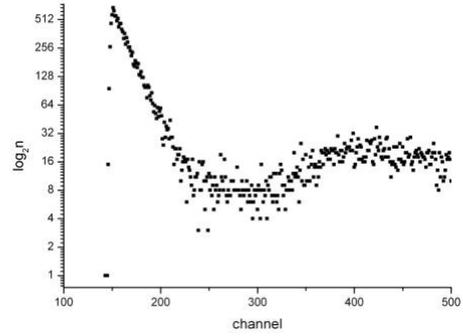

Fig.3. (b) Platform Enlarging

The measured reaction rates for the $^{239}$Pu fission chamber at four states are shown in Table 1.

Table 1. Reaction rates of the $^{239}$Pu fission chamber with different covers

| Fission chamber | $^{10}$B cover | Cadmium cover | Naked | Polythene cover |
|---|---|---|---|---|
| Reaction Rate/×10$^{-16}$s$^{-1}$ | 1.791 | 2.350 | 13.85 | 256.9 |

### 3.2 Discussions

The cadmium ratio of $^{239}$Pu in the fast neutron critical assembly is 5.89 indicating that the neutrons beyond 1eV contribute to 17.0% (1/5.89) of the reaction rate. The thermal neutron spectrum is usually treated as the 1/v spectrum. Then the $^{239}$Pu average cross-section for the thermal neutrons is calculated to be 500b based on this assumption. Then the thermal neutron flux is measured to be $2.30 \times 10^6$ s$^{-1} \cdot$ cm$^{-2}$.

The $^{10}$B cover absorbs the neutrons below 10keV. After the absorption of the resonance neutrons by the $^{10}$B cover, the effective cross-section of (n, f) drops. The $^{10}$B cover affects the fission rate of $^{239}$Pu mainly by absorbing the neutrons below 10keV. Using the calculated neutron spectrum, the cross-section of $^{239}$Pu (n, f) reaction and $^{10}$B (n, α) reaction from ENDF/B-VII.1 database, the average effect cross-section of $^{239}$Pu (n, f) reaction is to be 1.72b as neutron energy is beyond 10keV. Then the fast neutron flux beyond 10keV is $1.04 \times 10^8$ s$^{-1}$ · cm$^{-2}$.

The resonance neutron (energy between 1eV and 10keV) spectrum is generally using 1/E spectrum. Then the average cross-section in resonance energy region is calculated to be 24.95b. And the resonance region neutron flux can be determined to be $2.24 \times 10^6$ s$^{-1}$ · cm$^{-2}$. The total neutron flux is $1.09 \times 10^8$ s$^{-1}$ · cm$^{-2}$, with the thermal, resonance and fast neutrons are 2.1%, 2.1% and 95.8% respectively, yet the reaction contribution fraction are 83.0%, 4.1%, and 12.9% respectively.

Table 2. Neutron flux rate in different energy region

| Energy Region | Neutron Flux s$^{-1}$ cm$^{-2}$ | Fraction | Contribution to the Reaction Rate |
|---|---|---|---|
| <1eV | $2.30 \times 10^6$ | 2.1% | 83.0% |
| 1eV<En<10keV | $2.24 \times 10^6$ | 2.1% | 4.1% |
| >10keV | $1.04 \times 10^8$ | 95.8% | 12.9% |

As the quotient of thermal neutrons increasing, the reaction rate raises sharply. When the fission chamber is covered by a 2cm thick polythene, the reaction rate of $^{239}$Pu increases to $2.569 \times 10^{-14}$ s$^{-1}$, which is 18.5 times reaction rate of the bare $^{239}$Pu fission chamber. We calculate the neutron spectrum after the moderation by the 2cm thick polythene, getting that thermal, resonance and fast neutrons are 2.4%, 10.6% and 87.0% respectively. According to the total neutron flux the reaction rate is $1.757 \times 10^{-15}$s$^{-1}$, and thermal, resonance and fast neutrons contribution fraction are 74.4%, 16.3% and 9.3%. It is much smaller than $2.569 \times 10^{-14}$s$^{-1}$. Because the structural material around the assembly is not taken into consideration in the calculation, the thermal neutrons are zero. Actually, there are often a few thermal neutrons and resonance neutrons for the moderating of the structural material. The difference between the calculation and the experiment can also reveal the great impact of the thermal and resonance neutrons.

The $^{239}$Pu fission chamber is very sensitive to thermal and resonance neutrons. Therefore, the measurement for neutron flux of the slightly moderated fission spectrum by the $^{239}$Pu fission chamber can be divided into three sections: (a) use the bare $^{239}$Pu fission chamber to measure the total neutron reaction contribution; (b) use the $^{239}$Pu fission chamber covered by $^{10}$B to measure the fast neutron flux; (c) use the $^{239}$Pu fission chamber covered by Cd to measure reaction rate while obviating thermal neutrons. Finally the thermal, resonance and fast neutron flux can be measured respectively.

## 4  Conclusions

On the assumption that the resonance neutron spectrum and the thermal neutron spectrum are satisfying 1/E and 1/v distribution respectively, using a bare, a Cd-covered and a $^{10}$B-covered $^{239}$Pu fission chamber separately, the thermal, resonance and fast neutron flux can be measured. The thermal, resonance and fast neutron flux are achieved to be $2.30 \times 10^6$, $2.24 \times 10^6$ and $1.04 \times 10^8$ cm$^{-2}$ · s$^{-1}$ at a Godiva or Flattop type critical assembly; and the proportion of thermal, resonance and fast neutrons are 2.1%, 2.1%, and 95.8%; but the reaction contribution proportion are 83.0%, 4.1% and 12.9%.

Since the cross-section of thermal neutrons is much greater than that of the fast neutrons, a slightly change of thermal neutrons will lead to large variation in reaction rate. For this reason the neutrons of fast critical assemblies can be divided into three sections to measure the neutron flux by $^{239}$Pu fission chamber. The result shows that the bare, Cd-covered and $^{10}$B-covered fission chambers can be used to measure the three region neutron flux.

## References

[1] Yu V Stenkin, V V Alekseenko, D M Gromushkin, et. al. Thermal neutron flux produced by EAS at various altitudes[J]. Chinese physics C. 2013, 37(1): 015001-015001.

[2] WANG Wen-Xin, ZHANG Yi, WANG Ji-Jin, HU Bi-Tao. Study on spatial resolution of micromegas as a neutron detector under condition of high neutron flux and γ ray background[J]. Chinese physics C. 2009, 33(2): 110-113.

[3] O.Cabellos, P. Fernández, D. Rapisardac, N. García-Herranz et. al. Assessment of fissionable material behaviour in fission chambers[J]. Nuclear Instruments and Methods in Physics Research Section A. 2010, 618(1-3):248-259.

[4] SONG Yu-shou, Margaryan A., HU Bi-Tao, TANG Li-Guang. Performance of the low pressure MWPCs for fission fragments under a high background[J]. Chinese physics C. 2011,35(8): 758-762.

[5] S V Chuklyaev, Yu N Pepyolyshev, and V A Artem'ev. A Channel for Neutron Flux Measurement[J]. Instruments and Experimental Techniques. 2002, 45(2): 162-166.

[6] UNESAKI Hironobu, IWASAKI Tomohiko, KITADA Takanori, et. al. Measurement of $^{237}$Np fission rate ratio relative to 235U fission rate in cores with various thermal neutron spectrum at the Kyoto University Critical Assembly. [J]. Nucl. Sci. Technol. 2000, 37(8): 627-635.

[7] Loaiza D, Daniel Gehman. End of an Era for the Los Alamos Criticality Experiments Facility: History of Critical Assemblies and Experiments (1946-2004) [J]. Annals of Nuclear Energy. 2006, 33: 1339–1359.

[8] Improved design and construction of an ionization chamber for the CSNSbeam loss monitor (BLM) [J]. Chinese physics C. 2012, 36(4): 329-333.

[9] M. Calviani, P. Cennini, D. Karadimos, V. Ketlerov, V. Konovalov, W. Furman, A. Goverdowski, V. Vlachoudis, L. Zanini, the n_TOF Collaboration. A fast ionization chamber for fission cross-section measurements at n_TOF[J]. Nuclear Instruments and Methods in Physics Research Section A. 2008, 594(2): 220-227.

[10] P. Loiseau, et. al. On the fission chamber pulse charge acquisition and interpretation at MINERVE [J]. Nuclear Instruments and Methods in Physics Research Section A. 2013, 707: 58-63.